\documentclass[aip,jap,reprint,nofootinbib,superscriptaddress]{revtex4-1}

\usepackage{amsfonts} 
\usepackage{amssymb}
\usepackage{graphicx}
\usepackage{tabularx}
\usepackage{tabulary}
\usepackage{upgreek}
\usepackage{subfigure}
\usepackage{psfrag}
\usepackage{color}
\usepackage{braket}
\usepackage{natbib}

\definecolor{light-gray}{gray}{0.8}
\usepackage[left=1.3cm,top=2cm,bottom=2cm,right=1.1cm,foot=1.5cm]{geometry}

\begin{document}
\title{\huge{Order of Magnitude Smaller Limit on the Electric Dipole Moment of the Electron}}\vspace{-40pt}

\author{The ACME Collaboration$^{*}$: J. Baron$^1$\noaffiliation}
\author{W. C. Campbell$^2$\noaffiliation}
\author{D. DeMille$^3$\noaffiliation}
\author{J. M. Doyle$^1$\noaffiliation}
\author{G. Gabrielse$^1$\noaffiliation}
\author{Y. V. Gurevich$^{1,**}$\noaffiliation}
\author{P. W. Hess$^1$\noaffiliation}
\author{N. R. Hutzler$^1$\noaffiliation}
\author{E. Kirilov$^{3,\#}$\noaffiliation}
\author{I. Kozyryev$^{3,\dagger}$\noaffiliation}
\author{B. R. O'Leary$^3$\noaffiliation}
\author{C. D. Panda$^1$\noaffiliation}
\author{M. F. Parsons$^1$\noaffiliation}
\author{E. S. Petrik$^1$\noaffiliation}
\author{B. Spaun$^1$\noaffiliation}
\author{A. C. Vutha$^4$\noaffiliation}
\author{A. D. West$^3$\noaffiliation}
\maketitle

\newcommand{\Eeff}{\mathcal{E}_\mathrm{eff}}
\newcommand{\vecEeff}{\vec{\mathcal{E}}_\mathrm{eff}}
\newcommand{\de}{d_e} 

\newcommand{\vecde}{\vec{d}_e} 
\newcommand{\wNE}{\omega^{\mathcal{NE}}}

\newcommand{\N}{\mathcal{N}} 
\newcommand{\E}{\mathcal{E}} 
\newcommand{\B}{\mathcal{B}} 
\newcommand{\sign}{\textrm{sign}}
\newcommand{\A}{\mathcal{A}} 

\newcommand{\wNEresult}{$(2.6 \pm 4.8_\mathrm{stat} \pm 3.2_\mathrm{syst})$ mrad/s}
\newcommand{\result}{$(-2.1 \pm 3.7_\mathrm{stat} \pm 2.5_\mathrm{syst})\times 10^{-29}$ $e$ cm}
\newcommand{\upperlimit}{$8.7\times 10^{-29}$ $e$ cm}

\let\thefootnote\relax\footnotetext{$^1$Department of Physics, Harvard University, 17 Oxford Street, Cambridge, Massachusetts 02138, USA. $^2$Department of Physics and Astronomy, University of California Los Angeles, 475 Portola Plaza, Los Angeles, CA 90095, USA. $^3$Department of Physics, Yale University, 217 Prospect Street, New Haven, Connecticut 06511, USA. $^4$Department of Physics and Astronomy, York University, 4700 Keele Street, Toronto, Ontario M3J 1P3, Canada.$^{**}$Now at Department of Physics, Yale University, $^\#$ Now at Institut f\"{u}r Experimentalphysik, Universit\"{a}t Innsbruck, Technikerstrasse 25/4, A-6020 Innsbruck, Austria.$^\dagger$Now at Department of Physics, Harvard University. $^{*}$Correspondence and requests for materials should be addressed to acme@cua.harvard.edu.}

\small{\textbf{The Standard Model (SM) of particle physics fails to explain dark matter and why matter survived annihilation with antimatter following the Big Bang.  Extensions to the SM, such as weak-scale Supersymmetry, may explain one or both of these phenomena by positing the existence of new particles and interactions that are asymmetric under time-reversal (T). These theories nearly always predict a small, yet potentially measurable ($10^{-27}$-$10^{-30}$ $e$ cm) electron electric dipole moment (EDM, $\de$), which is an asymmetric charge distribution along the spin ($\vec{S}$). The EDM is also asymmetric under T. Using the polar molecule thorium monoxide (ThO), we measure $\de = $\result. This corresponds to an upper limit of $|\de| <$ \upperlimit$\:$with 90 percent confidence, an order of magnitude improvement in sensitivity compared to the previous best limits. Our result constrains T-violating physics at the TeV energy scale.}}


\small{
The exceptionally high internal effective electric field ($\Eeff$) of heavy neutral atoms and molecules can be used to precisely probe for $d_e$ via the energy shift $U = -\vecde\cdot\vecEeff $, where $\vec{d}_e=\de \vec{S}/(\hbar/2)$. Valence electrons travel relativistically near the heavy nucleus, making $\Eeff$ up to a million times larger than any static laboratory field\cite{Sandars1965,Khriplovich1997,Commins2010}. The previous best limits on $\de$  came from experiments with thallium (Tl) atoms\cite{Regan2002} ($|\de|<1.6\times 10^{-27}$ $e$ cm), and ytterbium fluoride (YbF) molecules\cite{Hudson2011,Kara2012} ($|\de|<1.06\times10^{-27}$ $e$ cm).
The latter demonstrated that molecules can be used to suppress the motional electric fields and geometric phases that limited the Tl measurement\cite{Hudson2011} (this suppression is also present in certain atoms\cite{Player1970}). Insofar as molecules can be fully polarized in laboratory-scale electric fields ($\E$), $\Eeff$ can be much greater than in atoms. The $^3\Delta_1$  electronic state used in ThO provides an $\Eeff\approx84$ GV/cm, the largest yet used in any EDM measurement\cite{Skripnikov2013,Meyer2008}. Its unusually small magnetic moment reduces its sensitivity to spurious magnetic fields\cite{Vutha2010,Vutha2011}. Improved systematic error rejection is possible because internal state selection allows the reversal of $\Eeff$ with no change in $\vec{\E}$\cite{Bickman2009,Eckel2013}.

To measure $\de$ we perform a spin precession measurement\cite{Vutha2010,Campbell2013,Kirilov2013} on a pulse of $^{232}$Th$^{16}$O molecules from a cryogenic buffer gas beam source\cite{Maxwell2005,Hutzler2011}. The pulse passes between parallel plates that generate a laboratory electric field $\E_z\hat{z}$ (Figure \ref{fig:apparatus_overview}). A coherent superposition of two spin states, corresponding to a spin aligned in the $xy$ plane, is prepared using optical pumping and state preparation lasers. Parallel electric ($\vec{\mathcal{E}}$) and magnetic ($\vec{\mathcal{B}}$) fields exert torques on the electric and magnetic dipole moments, causing the spin vector to precess in the  $xy$ plane. The precession angle is measured with a readout laser and fluorescence detection. A change in this angle as $\vecEeff$ is reversed is proportional to $d_e$.

\begin{figure}[!h]
\centering
\includegraphics[scale=0.47]{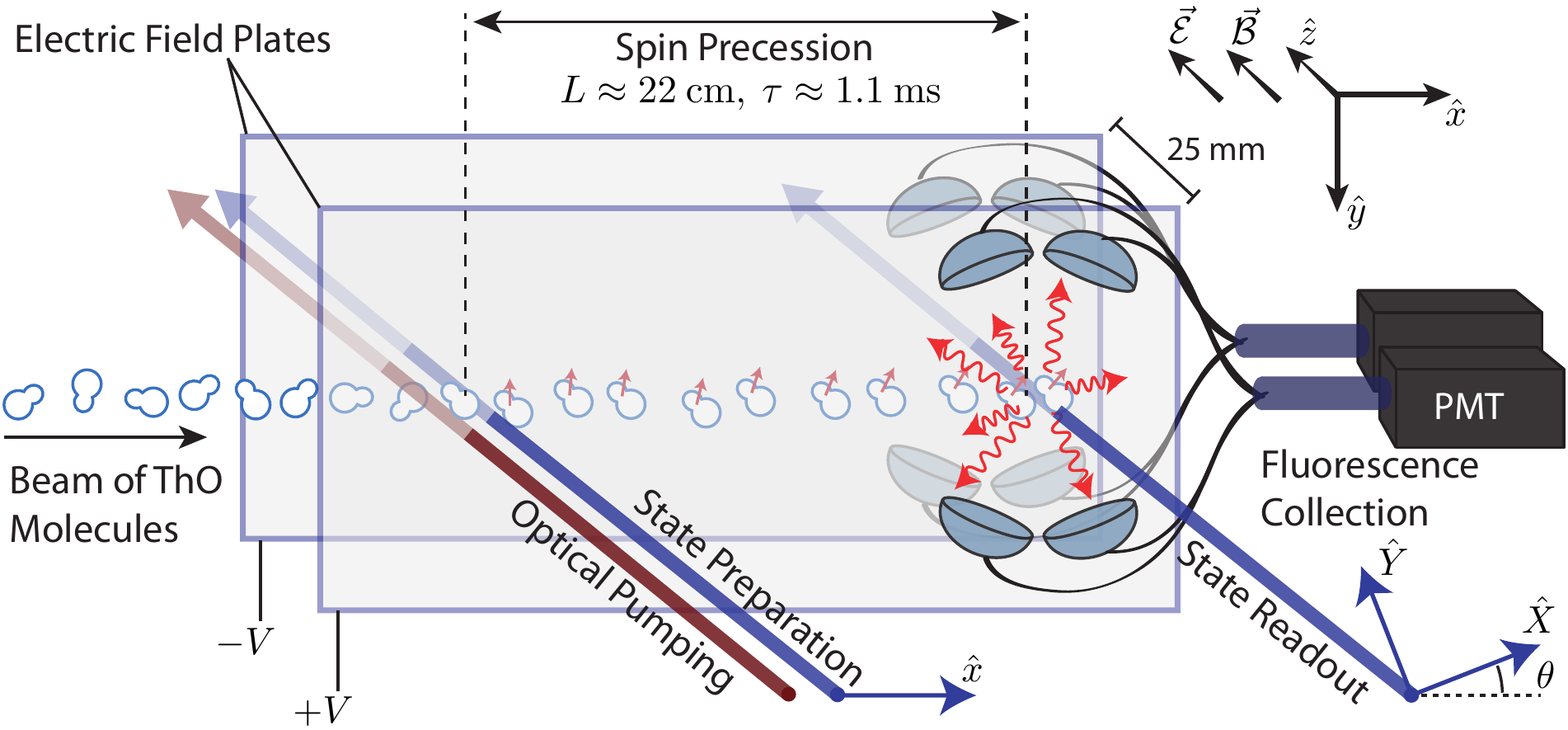}
\caption{\footnotesize{Schematic of the apparatus (not to scale). A collimated pulse of ThO molecules enters a magnetically shielded region. An aligned spin state (smallest red arrows), prepared via optical pumping, precesses in parallel electric and magnetic fields. The final spin alignment is read out by a laser with rapidly alternating linear polarizations, $\hat{X},\hat{Y}$, with the resulting fluorescence collected and detected with photomultiplier tubes (PMTs).}\label{fig:apparatus_overview}}
\end{figure}

\begin{figure}
\begin{centering}
\includegraphics[scale=.85]{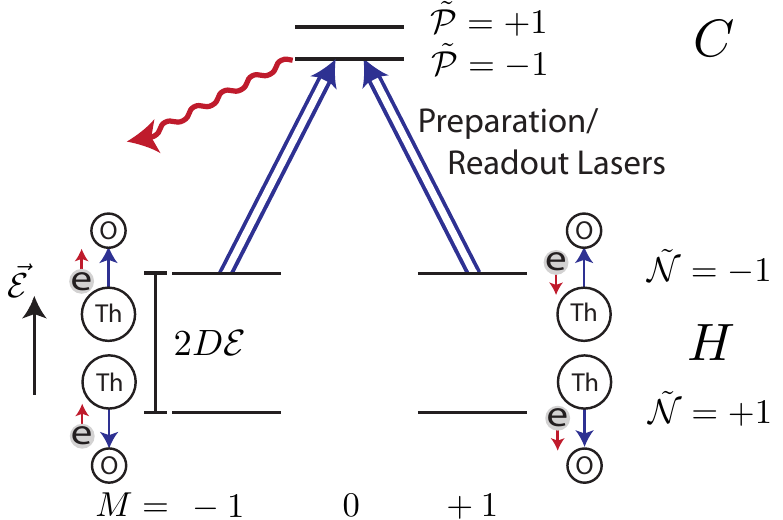}
\par\end{centering}

\caption{\footnotesize{Energy level diagram showing the relevant states. The state-preparation and readout
lasers (double lined blue arrows) drive one molecule orientation $\tilde{\mathcal{N}}=\pm1$
(split by $2D\mathcal{E}\sim100^{\:}\mathrm{MHz}$) in the $H$ state to $C$, with parity $\tilde{\mathcal{P}}=\pm1$ (split by $50^{\:}\mathrm{MHz}$). Population in the $C$ state
decays via spontaneous emission, and we detect the resulting fluorescence.
$H$ state levels are accompanied by cartoons displaying the
orientation of $\vec{\E}_\mathrm{eff}$ (blue arrows) and the spin
of the electron (red arrows) that dominantly contributes
to the $d_e$ shift.}}
\label{fig:H_C_Energy_Levels}
\end{figure}

In more detail, a 943 nm laser beam optically pumps molecules from the ground electronic state into the lowest rotational level, $J=1$, of the metastable (lifetime $\sim 2$ ms) electronic $H^{3}\Delta_{1}$ state manifold, in an incoherent mixture of the $\tilde{\mathcal{N}}=\pm1$, $^{\:}M=\pm1$ states. $M$ is the angular momentum projection along the $\hat{z}$ axis. 
$\tilde{\mathcal{N}}$ refers to the internuclear axis, $\hat{n}$, aligned ($+1$) or anti-aligned ($-1$) with respect to $\vec{\mathcal{E}}$, when $|\mathcal{E}|\gtrsim1^{\:}\mathrm{V/cm}$\cite{Vutha2011}. The linearly polarized state-preparation laser's frequency is resonant with the $H\rightarrow C$ transition at 1090 nm (Figure \ref{fig:H_C_Energy_Levels}). Within the short-lived ($\sim 500$ ns) electronic $C$ state there are two opposite parity $\tilde{\mathcal{P}}=\pm1$ states with $J=1,M=0$. For a given spin precession measurement, the laser frequency determines the $\tilde{\mathcal{N}},\tilde{\mathcal{P}}$ states that are addressed. This laser optically pumps the ``bright'' superposition of the two resonant $M=\pm1$ sublevels out of the $H$ state, leaving behind the ``dark'' orthogonal superposition that cannot absorb the laser light. If the state-preparation laser were polarized along $\hat{x}$, the prepared state, $|\psi(\tau=0),\tilde{\mathcal{N}}\rangle$, has the electron spin aligned along the $\hat{y}$ axis. The spin then precesses in the $xy$ plane by angle $\phi$ to
\begin{equation}
|\psi(\tau),\tilde{\mathcal{N}}\rangle=(e^{-i\phi}|M=+1,\tilde{\mathcal{N}}\rangle+e^{+i\phi}|M=-1,\tilde{\mathcal{N}}\rangle)/\sqrt{2}.
\end{equation}
As $\vec{\E}$ and $\vec{\B}$ are aligned along $\hat{z}$}, the phase $\phi$ is determined by $|\mathcal{B}_z| = |\vec{\mathcal{B}}\cdot\hat{z}|$,
its sign, $\tilde{\mathcal{B}}=\mathrm{sgn}(\vec{\mathcal{B}}\cdot\hat{z})$,
 and the electron's EDM, $d_e$:
\begin{equation}
\phi  \approx -(\mu_{\mathrm{B}}g\tilde{\mathcal{B}}\left|\mathcal{B}_{z}\right|+\tilde{\mathcal{N}}\tilde{\mathcal{E}}d_{e}\mathcal{E}_{\mathrm{eff}})\tau/\hbar,
\end{equation}
where $\tilde{\mathcal{E}}\equiv\mbox{sgn}(\vec{\mathcal{E}}\cdot\hat{z})$,
 $\mu_\mathrm{B}g$ is the magnetic moment\cite{Kirilov2013}, and $\tau$  is the spin precession time. The sign of the EDM term, $\tilde{\mathcal{N}}\tilde{\mathcal{E}}$, arises from the relative orientation between the $\vec{\E}_\mathrm{eff}$ and the electron spin, as illustrated in Figure \ref{fig:H_C_Energy_Levels}.
 
After the spin precesses over a distance of $L\approx22^{\:}\mathrm{cm}$ ($\tau\approx1.1^{\:}\mathrm{ms}$), we measure $\phi$ by optically pumping on the same $H\rightarrow C$ transition with the state readout laser. The laser polarization alternates between $\hat{X}$ and $\hat{Y}$ every 5 $\upmu$s, and we record the modulated fluorescence signals $S_X$ and $S_Y$ from the decay of $C$ to the ground state. This procedure amounts to a projective measurement of the spin onto $\hat{X}$ and $\hat{Y}$, which are defined such that $\hat{X}$ is at an angle $\theta$ with respect to $\hat{x}$ in the $xy$ plane. To normalize out molecule number fluctuations, we compute the asymmetry,\cite{Vutha2010}
\begin{equation}
\mathcal{A}\equiv\frac{S_{X}-S_{Y}}{S_{X}+S_{Y}}=\mathcal{C}\cos\left(2\left(\phi-\theta\right)\right)
\end{equation}
where the contrast $\mathcal{C}$ is $94\pm2\%$ on average. We set $\left|\mathcal{B}_{z}\right|$ and $\theta$ such that $\phi-\theta\approx\frac{\pi}{4}\left(2n+1\right)$ for integer $n$, so that the asymmetry is linearly proportional to small changes in $\phi$, and maximally sensitive to the EDM. We measure $\mathcal{C}$ by dithering $\theta$ between two nearby values that differ by $0.1$ rad, denoted by $\tilde{\theta} =\pm1$.

We perform this spin precession measurement repeatedly under varying experimental conditions to (a) distinguish the EDM energy shift from background phases and (b) search for and monitor possible systematic errors. Within a ``block" of data taken over 40 s, we perform measurements of the phase for each experimental state derived from 4 binary switches, listed from fastest (.5 s) to slowest (20 s): the molecule alignment, $\tilde{\mathcal{N}}$; the $\E$-field direction, $\tilde{\mathcal{E}}$; the readout laser polarization dither state, $\tilde{\theta}$; and the $\B$-field direction, $\tilde{\mathcal{B}}$.  For each $(\tilde{\mathcal{N}},\tilde{\mathcal{E}},\tilde{\mathcal{B}})$ state of the experiment, we measure $\mathcal{A}$ and $\mathcal{C}$, from which we can extract $\phi$. Within each block, we form ``switch-parity components" of the phase, $\phi^u$, that are combinations of the measured phases that are odd or even under these switch operations\cite{Eckel2013}. We denote the switch-parity of a quantity with a superscript, $u$, listing the switch labels under which the quantity is odd; it is even under all unlabeled switches. 
For example, the EDM contributes to a phase component $\phi^\mathcal{NE}=-d_e \Eeff \tau /\hbar$. 
We extract the mean precession time $\tau$ from $\phi^{\mathcal{B}}=-\mu_{\mathrm{B}}g\left|\mathcal{B}_{z}\right|\tau/\hbar$ and compute the frequencies, $\omega^{u}\equiv\phi^{u}/\tau$. The EDM value is obtained from $\wNE$ by $d_{e}=-\hbar\omega^{\mathcal{NE}}/\mathcal{E}_{\mathrm{eff}}$.

On a slower time scale, we perform additional ``superblock" binary switches to suppress some known systematic errors and to search for unknown ones. These switches, which occur on the 40--1200 s time scales, are: (1) the excited state parity addressed by the state readout lasers, $\tilde{\mathcal{P}}$; (2) a rotation of the readout polarization basis by $\theta\rightarrow\theta+\pi/2$, $\tilde{\mathcal{R}}$; (3) a reversal of the leads that supply the electric fields, $\tilde{\mathcal{L}}$; and (4) a global polarization rotation of both the state preparation and readout laser polarizations, $\tilde{\mathcal{G}}$. The $\tilde{\mathcal{P}}$ and $\tilde{\mathcal{R}}$ switches interchange the role of the $\hat{X}$ and $\hat{Y}$ readout beams and hence reject systematic errors associated with small differences in power, shape, or pointing. The two $\tilde{\mathcal{G}}$ state angles are chosen to suppress systematics that couple to unwanted ellipticity imprinted on the polarizations by birefringence in the electric field plates. The $\tilde{\mathcal{L}}$ switch rejects systematics that couple to an offset voltage in the electric field power supplies. We extract the EDM from $\wNE$ after a complete set of the $2^{8}$ block and superblock states. The $\wNE$ is even under all of the superblock switches.

\begin{figure}[!ht]
\centering
\includegraphics[scale=0.45]
{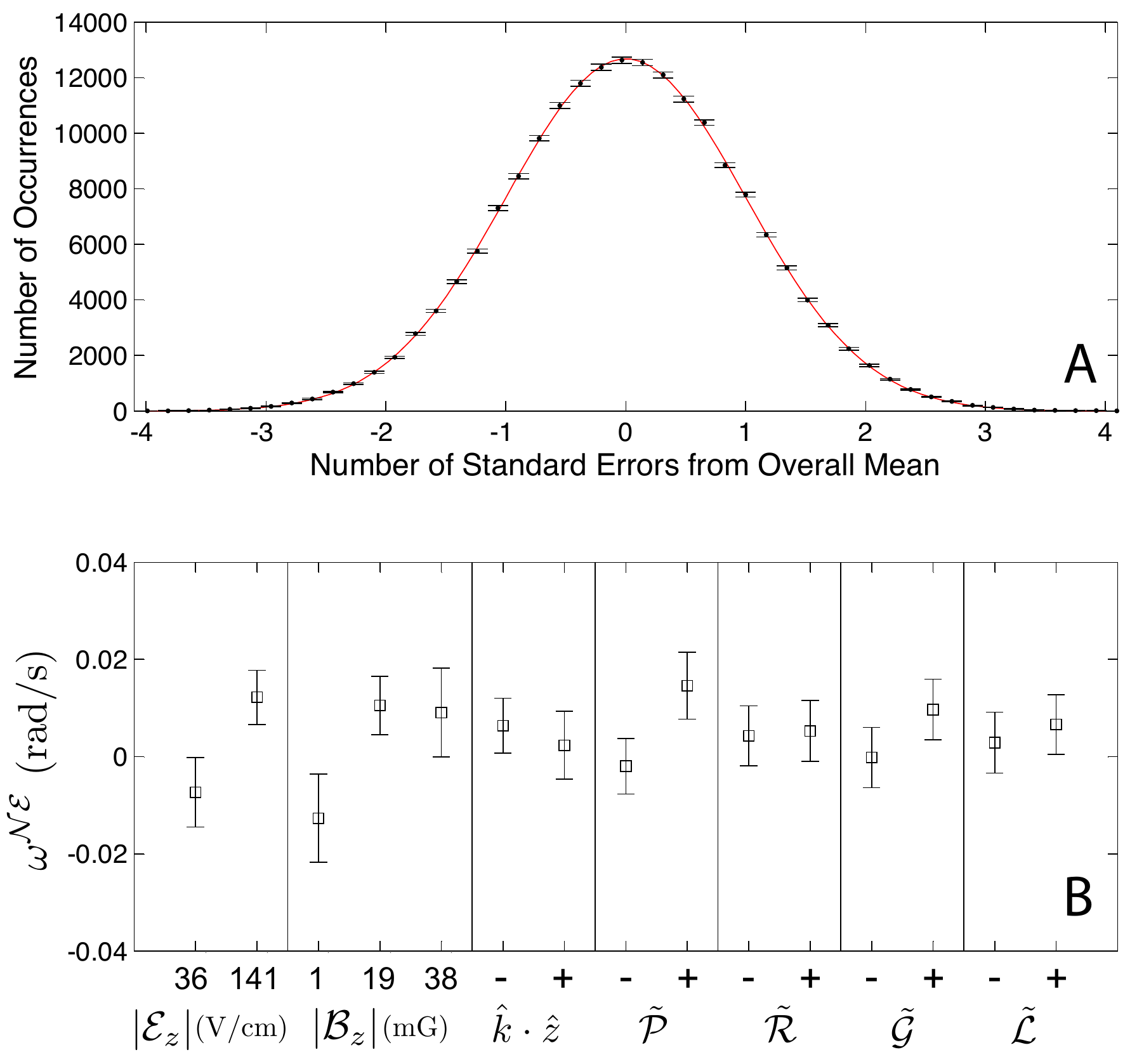}
\caption{\footnotesize{\textbf{(A)} Histogram of $\wNE$ measurements for each time point (within molecule pulse) and for all blocks. Error bars represent expected Poissonian fluctuations in each histogram bin. \textbf{(B)} Measured $\wNE$ values grouped by the states of $\left|\mathcal{B}_{z}\right|,$ $\left|\mathcal{E}_{z}\right|$, $\hat{k}\cdot\hat{z}$, and each superblock switch, before systematic corrections.\label{fig:edm_stats}}}
\end{figure}

The total dataset consists of $\sim 10^{4}$ blocks of data, taken over the course of $\sim$ 2 weeks. During this dataset, we also varied, from fastest (hours) to slowest (a few days): the $\mathcal{B}$-field magnitude, $\left|\mathcal{B}_{z}\right|\approx1,^{\:}19,^{\:}38^{\:}\mathrm{mG}$ (corresponding to $\left|\phi\right|\approx 0, \frac{\pi}{4}, \frac{\pi}{2}$ respectively), the $\mathcal{E}$-field magnitude $\left|\mathcal{E}_{z}\right|\approx36,^{\:}141^{\:}\mathrm{V/cm}$, and the pointing direction of the lasers, $\hat{k}\cdot\hat{z}=\pm1$. Figure \ref{fig:edm_stats}B shows measured EDM values obtained when the dataset is grouped according to the states of $\left|\mathcal{B}_{z}\right|,\left|\mathcal{E}_{z}\right|$, $\hat{k}\cdot\hat{z}$, and each superblock switch. All of these measurements are consistent within $2\sigma$.

We compute the standard error in the mean and use standard Gaussian error propagation to obtain the reported statistical uncertainty. The reported upper limit is computed using the Feldman-Cousins prescription\cite{Feldman1998} applied to a folded normal distribution. To prevent experimental bias, we performed a blind analysis by adding an unknown offset to $\wNE$.  The mean, statistical error, systematic shifts, and procedure for calculating the systematic error were determined before unblinding. Figure \ref{fig:edm_stats}A shows a histogram of EDM measurements. The asymmetry, $\mathcal{A}$, obeys a ratio distribution, which has large non-Gaussian tails in the limit of low signal to noise\cite{Curtiss1941}. We apply a photon count rate threshold cut so that we only include data with a large signal-to-noise, resulting in a statistical distribution that closely approximates a Gaussian. When the EDM measurements are fit to a constant value, the reduced chi-squared is $\chi^{2}=0.996\pm 0.006$. Based on the total number of detected photoelectrons ($\sim 1000$ per pulse) that contribute to the measurement, the statistical uncertainty is 1.15 times that from shot noise\cite{Kirilov2013}.

To search for possible sources of systematic error, we varied over 40 separate parameters and observed their effect on $\wNE$ and many other components of the phase correlated with $\tilde{\N}$,  $\tilde{\E}$, or $\tilde{\B}$. These parameters are intentionally applied tunable imperfections, such as transverse magnetic fields or laser detunings. These systematic checks were performed concurrently with the 8 block and superblock switches.

We assume that $\wNE$ depends linearly on each parameter $P$, so that the possible systematic shift and uncertainty of $\wNE$ is evaluated from the measured slope, $S = \partial \wNE/\partial P$, and the parameter value during normal operation (obtained from auxiliary measurements). If $S$ is not monitored throughout the data set, we do not apply a systematic correction but simply include the measured upper limit in our systematic error budget. Data taken with intentionally applied parameter imperfections is used only for determination of systematic shifts and uncertainties. Table 1 lists all contributions to our systematic error. 

We identified two parameters which systematically shift the value of $\wNE$ within our experimental resolution. Both parameters couple to the AC Stark shift induced by the lasers. The molecules are initially prepared in the dark state with a spin orientation dependent on the laser polarization. If there is a polarization gradient along the molecular beam propagation direction, the molecules acquire a small bright state amplitude. Away from the center of a Gaussian laser profile, the laser can be weak enough that the bright state amplitude is not rapidly pumped away; it acquires a phase relative to the dark state due to their mutual energy splitting, given by the AC Stark shift. An equivalent phase is acquired in the state readout laser. 
This effect changes the measured phase by $\phi_{\mathrm{AC}}(\Delta,\Omega_\mathrm{r})\approx (\alpha\Delta + \beta\Omega_\mathrm{r})$, where $\Delta$, $\Omega_\mathrm{r}$ are the detuning and Rabi frequency of the $H\rightarrow C$ resonance, respectively. The constants $\alpha$, $\beta$ are measured directly by varying $\Delta$ and $\Omega_\mathrm{r}$, and depend on the laser's spatial intensity and polarization profile. These measurements are in good agreement with our analytic and numerical models.

\begin{figure}
\begin{center}
\includegraphics[width=0.45\textwidth]
{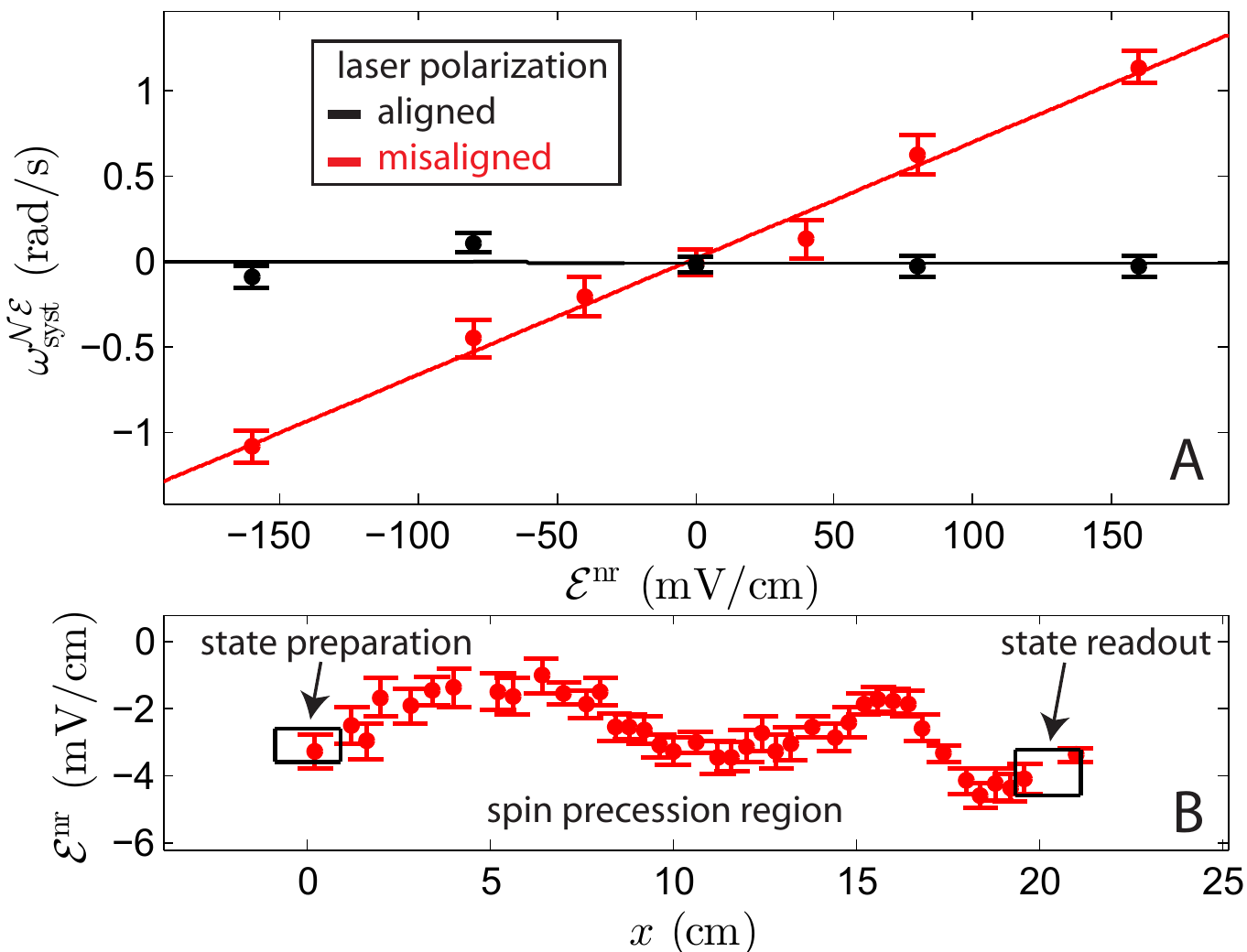}

\caption{
\footnotesize{\textbf{(A)} Tuning out laser polarization gradient and $\partial \wNE/\partial \mathcal{E}^\mathrm{nr}$ (see text for details). The red (black) points were taken with the polarization misaligned (aligned) with the birefringence axes of the electric field plates. 
\textbf{(B)} Microwave spectroscopic measurement of $\mathcal{E}^\mathrm{nr}$ along the molecule beam axis, $x$.
\label{fig:E_NR_Image}
\label{fig:enr_uwaves}}}
\end{center}
\end{figure}

A significant polarization gradient is caused by laser-induced thermal stress birefringence\cite{Eisenbach1992} in the electric field plates. The laser beams are elongated perpendicular to the molecular beam axis, which creates an asymmetric thermal gradient and defines the axes for the resulting birefringence gradient. By aligning the laser polarization with the birefringence axes, the polarization gradient can be minimized. We have verified this both with polarimetry\cite{Berry1977} and through the resulting AC Stark shift systematic (Figure \ref{fig:E_NR_Image}A).

Such AC Stark shift effects cause a systematic shift in our measurement of $\wNE$ in the presence of an $\tilde{\N}\tilde{\E}$ correlated detuning, $\Delta^{\N\E}$, or Rabi frequency, $\Omega_\mathrm{r}^{\N\E}$. We observe both.

The detuning component $\Delta^{\N\E}$ is caused by a non-reversing $\E$-field component $\E^\mathrm{nr}$, generated by patch potentials and technical voltage offsets, which is small relative to the reversing component, $|\E_z|\tilde{\E}$.
 The $\E^\mathrm{nr}$ creates a correlated DC Stark shift with an associated detuning $\Delta^{\mathcal{N}\mathcal{E}}=D \mathcal{E}^{\mathrm{nr}}$, where $D$ is the $H$ state electric dipole moment. We measured $\mathcal{E}^{\mathrm{nr}}$ via microwave spectroscopy (Figure \ref{fig:E_NR_Image}B), two-photon Raman spectroscopy, and by monitoring of the $\tilde{\mathcal{N}}\tilde{\mathcal{E}}$-correlated contrast.

The Rabi frequency component $\Omega_\mathrm{r}^{\mathcal{NE}}$, arises from a dependence of $\Omega_\mathrm{r}$ on the orientation of the molecular axis, $\hat{n}\approx\tilde{\mathcal{N}}\tilde{\mathcal{E}}\hat{z}$, with respect to laser propagation direction, $\hat{k}$. This $\hat{k}\cdot\hat{n}$ dependence can be caused by interference between E1 and M1 transition amplitudes on the $H \rightarrow C$ transition. Measurements of a non-zero $\tilde{\mathcal{N}}\tilde{\mathcal{E}}$-correlated fluorescence signal and an $\tilde{\mathcal{N}}\tilde{\mathcal{E}}\tilde{\mathcal{B}}$-correlated phase, both of which changed sign when we reversed $\hat{k}$, provided evidence for a nonzero $\Omega_\mathrm{r}^{\mathcal{NE}}$. These channels, along with their linear dependence on an artificial $\Omega^{\mathcal{NE}}_\mathrm{r}$ generated with an $\tilde{\mathcal{N}}\tilde{\mathcal{E}}$ correlated laser intensity, allowed us to measure $\Omega_\mathrm{r}^{\mathcal{NE}}/\Omega_{\mathrm{r}} = (-8.0 \pm 0.8)\times 10^{-3} (\hat{k}\cdot \hat{z})$, where $\Omega_{\mathrm{r}}$ is the uncorrelated (mean) Rabi frequency.

By intentionally exaggerating these parameters we verified that both $\mathcal{E}^{\mathrm{nr}}$ and $\Omega_\mathrm{r}^{\mathcal{NE}}$ couple to AC Stark shift effects to produce a false EDM. Figure \ref{fig:E_NR_Image}A illustrates our ability to suppress the measured $\wNE$ shift as a function of applied $\E^\mathrm{nr}$. The correlations $\partial \wNE/\partial\mathcal{E}^\mathrm{nr}$ and $\partial \wNE/\partial\Omega_\mathrm{r}^\mathcal{NE}$ were monitored at regular intervals throughout the data set. The resulting systematic corrections to $\wNE$ were all $< 1$ mrad/s.

For a subset of our data, the $\tilde{\N}$-correlated phase $\phi^\N$ was non-zero and drifted with time.  We identified the cause of this behavior as an $\tilde{\N}$-correlated laser pointing $\hat{k}^\N\cdot\hat{x}\approx 5$ $\mu$rad present in our optical frequency switching setup.  We eliminated this effect with improved optical alignment; however, since we were not able to determine the precise mechanism by which $\hat{k}^\N$ coupled to $\phi^\N$, we chose to include $\phi^\N$ variations in our systematic error budget. The slope $\partial\wNE/\partial\phi^\N$ (consistent with zero) and the mean value of $\phi^\N$ established a systematic uncertainty limit of $\approx$1 mrad/s on $\wNE$.

To be cautious, we include in our systematic error budget possible contributions from the following parameters that caused a non-zero EDM shift in experiments similar to ours: stray $\B$-fields $\B^{\mathrm{nr}}_{x,y,z}$ and $\B$-field gradients\cite{Eckel2013}; an $\tilde{\E}$-correlated phase, $\phi^\E$, caused by leakage current, $\vec{v} \times \vec{\E}$, and geometric phase effects\cite{Regan2002}; and laser detunings and $\E$-field ground offsets\cite{Hudson2011}.  We obtained direct $\wNE$ systematic limits of $\lesssim 1$ mrad/s for each. We simulated the effects that contribute to $\phi^\E$, by correlating $\B_z$ with $\tilde{\E}$, which allowed us to place a $\sim 10^{-2}$ mrad/s limit on their combined effect. Because of our slow molecular beam, relatively small applied $\E$-fields, and small magnetic dipole moment, we do not expect any of these effects to systematically shift $\wNE$ above the $10^{-3}$ mrad/s level\cite{Vutha2010,Vutha2011}.

\begin{table} \footnotesize
\begin{center}
\begin{tabular}{lcc}
\hline 
Parameter & Shift & Uncertainty\tabularnewline
\hline
\hline 
$\mathcal{E}^{\mathrm{nr}}$ correction & $-0.81$ & $0.66$\tabularnewline
$\Omega_{\mathrm{r}}^{\mathcal{NE}}$ correction & $-0.03$ & $1.58$\tabularnewline
$\phi^{\mathcal{E}}$ correlated effects & $-0.01$ & $0.01$\tabularnewline
$\phi^{\mathcal{N}}$ correlation &  & $1.25$\tabularnewline
Non-Reversing $\mathcal{B}$-field $\left(\mathcal{B}_{z}^{\mathrm{nr}}\right)$ &  & $0.86$\tabularnewline
Transverse $\mathcal{B}$-fields $\left(\mathcal{B}_{x}^{\mathrm{nr}},\mathcal{B}_{y}^{\mathrm{nr}}\right)$ &  & $0.85$\tabularnewline
$\mathcal{B}$-Field Gradients &  & $1.24$\tabularnewline
Prep./Read Laser Detunings &  & $1.31$\tabularnewline
$\tilde{\mathcal{N}}$ Correlated Detuning &  & $0.90$\tabularnewline
$\mathcal{E}$-field Ground Offset &  & $0.16$\tabularnewline
\hline 
Total Systematic & $-0.85$ & $3.24$\tabularnewline
\hline 
Statistical &  & $4.80$\tabularnewline
\hline
\hline 
Total Uncertainty &  & $5.79$\tabularnewline
\hline
\end{tabular}
\par\end{center}
\caption{\footnotesize{Systematic and statistical errors for $\wNE$, in units of mrad/s. All errors are added in quadrature. In EDM units, 1.3 mrad/s $\approx 10^{-29}$ $e$ cm.}}
\end{table}

The result of this first-generation ThO measurement,
\begin{equation}
\de = \textrm{\result,}
\end{equation}
comes from $d_{e}=-\hbar\omega^{\mathcal{NE}}/\mathcal{E}_{\mathrm{eff}}$ using $\Eeff=84$ GV/cm\cite{Skripnikov2013, Meyer2008} and $\wNE=$\wNEresult. This sets a 90 percent confidence limit,
\begin{equation}
|d_e| < \textrm{\upperlimit ,}
\end{equation}
that is 12 times smaller than the previous best limit\cite{Hudson2011, Regan2002}, an improvement made possible by the first use of the ThO molecule and of a cryogenic source of cold molecules for this purpose. Because paramagnetic molecules are sensitive to multiple T-violating effects\cite{Kozlov1995}, our measurement should be interpreted as $\hbar\wNE = -\de\Eeff - W_S C_S$, where $C_S$ is a T-violating electron-nucleon coupling, and $W_S$ is a molecule-specific constant\cite{Skripnikov2013,Dzuba2011a}. We assume $C_S=0$ for the $d_e$ limit above.  Assuming instead that $\de=0$ yields $C_S = (-1.3 \pm 3.0) \times 10^{-9}$, corresponding to a 90 percent confidence limit $|C_S| < 5.9 \times 10^{-9}$ that is 9 times smaller than the previous limit\cite{Griffith2009b}.

A measurably large EDM requires new mechanisms for T violation, equivalent to charge conjugation-parity (CP) violation, given the CPT invariance theorem\cite{Khriplovich1997}. Nearly every extension to the SM\cite{Barr1993, Pr05} introduces new CP violating phases $\phi_{\mathrm{CP}}$. It is difficult to construct mechanisms that systematically suppress $\phi_{\mathrm{CP}}$, so model builders typically assume $\sin(\phi_{\mathrm{CP}}) \sim 1$\cite{Engel2013}. An EDM arising from new particles at energy $\Lambda$ in an $n$-loop Feynman diagram will have size $d_e/e \!\sim\! \kappa(\alpha_{\mathrm{eff}}/4\pi)^n (m_e c^2/\Lambda^2)\mathrm{sin}(\phi_\mathrm{CP})(\hbar c)^{-1}$, where $\alpha_{\mathrm{eff}}\sim 4/137$ (for electroweak interactions) encodes the strength with which the electron couples to the new particles, $m_e$ is the electron mass, and $\kappa\sim$ 0.1 - 1 is a dimensionless prefactor\cite{Khriplovich1997,Fortson2003,Bernreuther1991}. In models where 1- or 2-loop diagrams produce $d_e$, our result typically sets a bound on CP violation at energy scales $\Lambda \sim 3$ TeV or $1$ TeV, respectively\cite{Barr1993,Pr05,Engel2013,Bernreuther1991}. Hence, within the context of many models, our more precise EDM limit constrains CP violation up to energy scales similar to or higher than those explored directly at the Large Hadron Collider.

\bibliography{references}

\bigskip

\noindent\textbf{Acknowledgements} This research was supported by NSF and the NIST PMG program.  We thank M. Reece and M. Schwartz for discussions, and S. Cotreau, J. MacArthur, and S. Sansone for technical support.
\bigskip

\noindent\textbf{Author Information} The authors declare no competing financial interests.

\newpage
\subsection*{Supplementary Materials}

We create a pulsed molecular beam of ThO using the buffer gas beam technique\cite{Hutzler2012,Hutzler2011,Maxwell2005}.  Each packet of molecules leaving the source contains $\sim 10^{11}$ ThO molecules in the $J=1$ rotational level of the ground electronic ($X$) and vibrational states and are produced at a repetition rate of 50 Hz.  The packet is 2-3 ms wide and has a center of mass speed of $\sim 200$ m/s.  After leaving the cryogenic beam source chamber, the molecules travel through a microwave field resonant with the $\ket{X;J=1}\leftrightarrow \ket{X;J=0}$ transition and optical pumping lasers resonant with the $\ket{X;J=2,3} \rightarrow \ket{C;J=1,2}$ transitions. The microwaves and optical pumping lasers transfer population from $\ket{X;J=0,2,3}$ into the $\ket{X;J=1}$ state leading to a twofold increase in its population. The molecules then pass through adjustable and fixed collimating apertures before entering the magnetically shielded interaction region, where electric and magnetic fields are applied. A retroreflected 943 nm laser optically pumps population from the $\ket{X;J=1,M=\pm 1}$ states to $\ket{A;J=0,M=0}$, which decays partially into the $\ket{H;J=1}$ state in which the EDM measurement is performed.

The spin precession region contains applied electric and magnetic fields, along with lasers to prepare and read our EDM state.  The electric field is provided by two plates of 12.7 mm thick glass coated with a layer of indium tin oxide (ITO) on one side, and an anti-reflection coating on the other.  The ITO coated sides of the plates face each other with a gap of 25 mm, and a voltage is applied to the ITO to create a uniform electric field.

The spatial profile of the electric field was measured by performing microwave spectroscopy on the ThO molecules.  When the molecule pulse is between the state preparation and read-out regions, a $40$ $\upmu$s burst of microwaves resonant with the DC Stark-shifted $\ket{H;J=1,M=\pm 1} \rightarrow \ket{H;J=2,M=0}$ transitions is introduced by a microwave horn at the end of the apparatus, counterpropagating to the molecular beam.  If on resonance, the microwaves drive a transition that spin-polarizes the molecules, similar to the state preparation scheme.  We can then detect the spin polarization using the normal readout scheme. The microwave transition width is $\sim 5$ kHz (dominated by Doppler broadening), so the $H$-state dipole moment of $D\approx 1$ MHz/(V/cm)\cite{Vutha2011} (for $J=1$) means that this method is sensitive to $\sim$ mV/cm electric field deviations with spatial resolution of $\approx 1$ cm, limited by the velocity distribution in the beam.  Our measurement indicated that the spatial variation of the electric field plate separation is $\sim 20\:\upmu$m across the molecule precession region, in very good agreement with an interferometric measurement\cite{Patten1971}. We can also test how well the electric field reverses by mapping the field with equal and opposite voltages on the plates.  This measurement indicated that the non-reversing component of the electric field had magnitude $|\mathcal{E}^\mathrm{nr}|\approx$ 1-5 mV/cm across the entire molecular precession region, as shown in Figure \ref{fig:enr_uwaves}B. 

The EDM measurement is performed in a vacuum chamber surrounded by five layers of mu-metal shielding.
The applied magnetic field is supplied by a cosine-theta coil, with several shim coils to create a more uniform magnetic field within the precession region, and to allow us to apply transverse magnetic fields and gradients for systematic checks.  Changes in the magnetic field are monitored by four 3-axis fluxgate magnetometers inside the magnetic shields, and the magnetic fields were mapped out before and after the experimental dataset was taken by sliding a 3-axis fluxgate down the beamline.

The lasers travel through the electric field plates, so all stages of the spin precession measurement are performed inside the uniform electric field.  All laser light in the experiment originates from external cavity diode lasers (ECDL), frequency stabilized via an Invar transfer cavity to a CW Nd:YAG laser locked to a molecular iodine transition\cite{Hall1999}.  All required transition frequencies and state assignments were determined previously\cite{Edvinsson1968,Edvinsson1990,Paulovic2003}.  We measured the saturation intensities, radiative lifetimes, electric/magnetic dipole moments, and branching ratios for all required states and transitions.

In order to normalize against drifting molecular beam properties (pulse shape, total molecule number, velocity mean and distribution, etc.), we perform a spin precession measurement every $10$ $\upmu$s, which is much faster than the molecular beam variations\cite{Kirilov2013}, spin precession time, and temporal width of the molecular pulse.  This is accomplished by sending the detection laser through two different beam paths, combined on the two ports of a polarizing beamsplitter.  The two beam paths can be rapidly switched on and off with acousto-optic modulators.

The transparent electric field plates allow us to collect a large fraction of the solid angle of fluorescence from the molecules.  Fluorescence travels through the field plates into an eight-lens system (four behind each plate) which focuses the light into an optical fiber bundle. The four bundles on each side are coupled into a fused quartz light pipe, which carries the fluorescence to a PMT (outside the magnetic shields).  The net detection efficiency, including collection solid angle and detector quantum efficiency, is $\approx 1\%$.  We typically register $\approx$ 1000 photon counts per molecule pulse.  The PMT photocurrents are read as analog signals by a low-noise, high-bandwidth amplifier, and then sent to a 24-bit digitizer operating at 5 megasamples/s.  The control and timing for all experimental parameters is managed by a single computer, and the timing jitter is less than one digitizer sampling period.

\end{document}